\documentclass[published]{JHEP3} 

\JHEP{00(2012)000}

\JHEPspecialurl{http://jhep.sissa.it/JOURNAL/JHEP3.tar.gz}

\usepackage{epsfig,multicol,bbm}

\newcommand\fverb{\setbox\fverbbox=\hbox\bgroup\verb}
\newcommand\fverbdo{\egroup\medskip\noindent%
			\fbox{\unhbox\fverbbox}\ }
\newcommand\fverbit{\egroup\item[\fbox{\unhbox\fverbbox}]}
\newbox\fverbbox

\title{Direct recovery of density fluctuation spectra from tomographic
  shear spectra}

\author{Marino Mezzetti$^{1,2}$, Silvio A. Bonometto$^{1,2,3}$,
  Luciano Casarini$^{4}$, Giuseppe Murante$^{1,5}$ \\ $^1$ --
  Department of Physics, Astronomy Unit, Trieste University, Via
  Tiepolo 11, I~34143 Trieste, Italy \\ $^2$ -- I.N.A.F. --
  Astronomical Observatory of Trieste, Via Tiepolo 11, I~34143
  Trieste, Italy \\ $^3$ -- I.N.F.N. -- Sezione di Trieste, Via
  Valerio, 2 I~34127 Trieste, Italy \\ $^4$ -- Departamento de Fisica,
  UFES, Avenida Fernando Ferrari 514, Vit\'oria, Esp\'\i rito Santo,
  Brasil \\ $^5$ -- I.N.A.F. -- Astronomical Observatory of Torino,
  Strada Osservatorio 20, I~10025 Pino Torinese (Torino), Italy \\ }

\received{.....}
\accepted{.....}

\abstract{Forthcoming experiments will enable us to determine high
  precision tomographic shear spectra. Matter density fluctuation
  spectra, at various $z$, should then be recovered from them, in
  order to constrain the model and determine the DE state equation.
  Available analytical expressions, however, do the opposite, enabling
  us to derive shear spectra from fluctuation spectra. Here we find
  the inverse expression, yielding density fluctuation spectra from
  observational tomographic shear spectra. The procedure involves
  $SVD$ techniques for matrix inversion. We show in detail how the
  approach works and provide a few examples.}


\begin{document}

\section{Introduction}
Dark Energy is the main finding and puzzle in today's cosmology. Its
contribution to the cosmic budget is directly constrained by Cosmic
Microwave Background spectra, but only measures of the matter density
field $\rho({\bf x},z)$, at low $z$, can provide clues on its state
equation $w(z)$. It is then important that weak lensing data, directly
sensitive to the whole matter distribution, can be translated into
information on the density fluctuation spectrum
\begin{equation}
\label{pk}
P(k,z) = \langle | \delta(k,z) |^2 \rangle
\end{equation}
and, namely, on its redshift ($z$) dependence. Here $\delta(k,z)$ is
the Fourier transform of the matter fluctuation field $\epsilon({\bf
  x},z) = \rho({\bf x},z)/\bar \rho - 1$. The final target would be
recovering $P(k,z)$ from tomographic shear data.

As a matter of fact, data can be used to obtain the tomographic shear
spectra $C_{ij}(\ell)$ --~defined below~-- and a known relation yields
$C_{ij}(\ell)$ from $P(k,z)$. In this paper we therefore aim at
inverting such a relation, so obtaining $P(k,z)$ from~$C_{ij}(\ell)$.

The problem somehow reminds the inversion of the Limber equation (see,
e.g., \cite{peebles}), aiming to obtain the angular 3-point
correlation function $\xi(r)$ (making no ansatz on its form) from the
2-point spatial function $w(\theta)$. There are quite a few
differences, however. In particular, 3--D galaxy catalogs made the
Limber equation obsolete. On the contrary, even having spectroscopic
redshifts for all lensed galaxies, no 3--D shear spectrum is recovered,
although the 2--D shear spectrum would then be known with great
accuracy. Furthermore, in order to recover fluctuation spectra, shear
spectra at various redshifts are to be simultaneously used; on the
contrary, the inversion of the Limber equation, even in the
relativistic regime \cite{BL}, is independently effective at various
redshifts.

Many authors debated the use of tomographic shear spectra to constrain
the cosmological model, with different procedures \cite{semboloni,kit}
(see also \cite{simon}). Also the Dark Energy task force (DEFT:
\cite{albrecht}) devoted much attention to this approach. The basic
pattern essentially amounts to comparing observational $C_{ij}(\ell)$
data with the theoretical $C_{ij}(\ell)$ obtainable for models
belonging to an assigned parameter space. This Bayesian procedure has
been used in quite a few former cases and is certainly effective,
while the technique discussed here is not yet mature enough to compete
with it. However, we see significant possibilities to upgrade it, to
at least obtain a complementary tool.

At present, significant shear data are already available. Cosmic shear
measurements were obtained by using large area ground surveys (see,
e.g., \cite{hoekstraFu}) or narrower area space data, characterized by
high quality imaging (see, e.g., \cite{schrabback}). However,
observational campaigns to perform ultimate systematic mapping of
tomographic cosmic shear are the basic aim of future missions.

In particular, the Euclid project, a recently approved ESA mission, is
devised to observe about half extra--galactic sky ($\sim 15,000$
  deg$^2$) from space, at a diffraction limited spatial resolution
which would be impossible from ground \cite{laureijs}. Euclid will
also obtain medium resolution (R~400) spectra of ~1/3 of all galaxies
brighter than 22 mag, in a wavelength range unreachable from ground
for faint galaxies above z=1.  By measuring the correlations in the
shapes of $\sim 1.5$ billion galaxies ($>2$ orders of magnitude more
than all galaxies in today's samples), Euclid will map weak
gravitational lensing with extreme accuracy, yielding $C_{ij}(\ell)$
with a precision $\cal O$$(1\, \%)$ {up to $\ell \sim 5000$. At
  larger $\ell$ (up to $\sim 30,000$) the precision gradually
  worsens.

At low $\ell$, cosmic shear can be derived from linear fluctuations.
Already at $\ell \sim 100$, however, non linear contributions exceed
$\sim 1\, \%$ while, above $\ell \simeq 500$, neglecting non linear
structures is clearly misleading. Baryonic physics causes spectral
shifts already at $\ell \sim 300$; they become more and more relevant
at larger $\ell$ and, above $\ell \simeq 2000$, shear spectra are
unpredictable if we neglect it.

According to Huterer \& Takada \cite{huterer}, data may become 
precise enough to
enable us to appreciate cosmological parameter variations causing a
shift $\cal O$$(1\, \%)$ in $P(k,z)$. This conclusion was attained by
parameterizing deviations in $P(k,z)$ and testing their consequences
on the angular shear spectra, trying also to take into account all
possible sources of systematic bias, but assuming that the deviations
in the shear spectra are solely due to the shift in the density power
spectrum $P(k,z)$.

In our case, while the power spectrum and its evolution measure the
linear and non--linear growth factors $\cal G$$(k,z)$, the relation
between $P(k,z)$ and the shear spectra depends on a {\it kernel}
involving a number of astrophysical assumptions (concerning, e.g., the
galaxy number distribution as a function of $z$ or the relation
between photometric and physical redshift) and a specific cosmological
assumption, the time dependence of the scale factor~$a$~(or the
redshift $z$).

This is why a parameter shift affecting the growth factor $\cal
G$$(k,z)$ seldom leaves the kernel unaffected. It may well be that the
kernel variations add up to $\cal G$$(k,z)$ variations, so
strenthening the Huterer--Takada effect. The opposite case is however
also possible.  If mutual cancellations occur, when some specific
parameter shift is considered, such specific parameter cannot be fixed
with the claimed accuracy. This is not just a theoretical {\it
  caveat}, as a number of examples can be given.

We stress this point also because a similar difficulty affects the
inversion procedure described below. When supposing that the shear
spectra are measured, we shall aim at reconstructing $P(k,z)$ -- and
thence the growth factor $\cal G$$(k,z)$ -- from them, by following a
direct analytical pattern. Clearly, two distinct time dependences
arise from the model choice: (i) background equations rule the
dependence of the scale factor $a$ on time and, therefore, space time
geometry; (ii) fluctuation dynamics rules the time dependence of the
growth factor $\cal G$. To achieve our {\it dynamical} aim, the
procedure we describe will assume that the {\it geometrical} kernel is
assigned.

In the discussion Section, we shall however return to this point. As a
matter of fact, geometry depends only on a part of the parameters
defining the model, while dynamical data open a window, e.g., on the
separate contributions of baryons and Dark Matter to cosmic matter and
on further parameters unrelated to the background metric, but critical
to define primeval fluctuations and their later evolution.
Furthermore, in the discussion Section we shall conjecture that this
apparent difficulty might be turned into a tool for model
discrimination.

The plan of the paper is as follows: In the next Section we shall
discuss the expression yielding $C_{ij}(\ell)$ from $P(k,z)$; this
will enable us to outline how such an expression can be formally
inverted. In Section 3 we shall enter into technical details
concerning the inversion procedure. In particular we shall introduce
the $SVD$ technique, essential for dealing with (nearly--)singular
matrices. The technique will enable us to provide a concrete solution
to the inversion problem, attaining a precision $\cal O$$(1:1000)$, at
least, that we shall illustrate by using Halofit fluctuation
spectra. In the same Section, however, a number of difficulties will
also be outlined. In Section 4 we shall go beyond Halofit, applying
the technique to hydrodynamical simulation outputs; we shall also show
an approach enabling us to overcome some of the difficulties
previously outlined. Finally, Section 5 is devoted to a discussion of
the results obtained and the perspectives opened.

\section{From fluctuation to shear spectra and viceversa}
The convergence weak lensing power spectra are linear functionals of
matter power spectra at various $z$, suitably convoluted with the
lensing properties of space, mostly due to the matter distribution in
it, and the background galaxy distribution. We set the galaxies, whose
images can be distorted by gravitational lensing, into $n$ bins at
increasing depth, labeled by $i,j=1,...,n~.$ Their distributions will
limit the functions $W_i(u)$, gauging the effects of the lensing
systems, in the expressions
\begin{equation}
C_{ij}(\ell) = H_{0}^{~4} \int_{0}^{\tau_0 } du~ W_{i}(u) W_{j}(u)~ P(\ell/u,u)
\label{pijl}
\end{equation}
yielding the tomographic shear spectra $C_{ij}(\ell) $ \cite{hu,
  refre, casarini2011}.  Here $\tau_0$ is the conformal age of the
Universe, $\tau = \tau_0 - u\, \, $ being the conformal time in the
FRW metric
\begin{equation}
ds^2 = a^2(\tau) \left( d\tau^2-d\lambda^2 \right)~,
\end{equation}
so that $d \lambda^{\, 2}$ is the co-moving 3--space metric, that we
assume to be flat; $P(k,u)$ is the fluctuation spectrum at the
conformal time set by $u$; $H_0$ is the Hubble parameter. The window
functions $W_i$ are then defined in the next subsection.

\subsection{Window functions}
In the literature, $n=1$, 3 or 5 bins were considered.  Data now
  available could not be analysed with $>5$ bins. With the use of the
  ordinary bayesian procedure, in fact, shot noise in data allows us
  no improvement in parameter determination already when going from 3
  to 5 bins (see, e.g., \cite{casarini2011}). As a matter of fact,
  shot noise adds to the $C_{ij}(\ell)$ spectra in a way $\propto
  n/N_g$ ($N_g:$ total number of lensed galaxies observed) and, above
  $n=3$, its effects confuse the signal dependence on the band
  ($i,j$), namely because of the approximation implicit in using
  photometric redshifts. However, if $N_g$ increases by a factor 100,
  as in future Euclid data, even with 10 bins shot noise can be
  ignored.  Here we shall however keep to a 5--bin case, to prevent
numerical complications in matrix algebra.

The bin limits $z_i$ are conveniently selected so to have the same
number of galaxies per bin; we shall also assume, as usual, that the
distribution of the galaxy number in redshift and solid angle reads
\begin{equation}
n(z) = {d^2 N \over d\Omega\, dz} = {\cal C} ~\bigg({z \over z_0
}\bigg)^A \exp\bigg[- \left( z \over z_0 \right)^B \bigg]
\end{equation}
with
\begin{equation}
{\cal C} = {B \over \left[z_0 \Gamma \left( A+1 \over B
\right) \right]}
\label{nz}
\end{equation}
and $A=2$, $B=1.5$, so that ${\cal C} = 1.5/z_0$ (with $z_0 =
z_m/1.412$ obtained from the median redshift $z_m = 0.9~$).

This distribution is then considered within the limits of the redshift
bins, taking however into account that only photometric redshift
values are given. The discrepancies between them and the actual galaxy
redshift define the filters
$$ 
\Pi_i (z) = \int_{z_{ph,i}}^{z_{ph,i+1}} dz' ~ {1 \over \sqrt{2 \pi}~
\sigma(z)} \exp \left(-{(z - z')^2 \over 2 \sigma^2 (z)} \right) =
~~~~~~~~~~~~
$$
\begin{equation}
~~~~~~~~ = {1 \over 2}
\left[
{\rm Erf}\left(z_{ph,i+1}-z \over \sqrt{2}\sigma(z)
\right)-{\rm Erf}\left(z_{ph,i}-z \over \sqrt{2}\sigma(z)\right)
\right]
\end{equation}
with $\sigma(z) = 0.05~(1+z)$ coherently with Euclid expectations
\cite{laureijs} (see also \cite{amaraAmendola}) and set
\begin{equation}
D_i(z) = n(z) \Pi_i(z)
\label{diz}
\end{equation}
\begin{figure}
\begin{center}
\vskip -.7truecm
\includegraphics[scale=0.45]{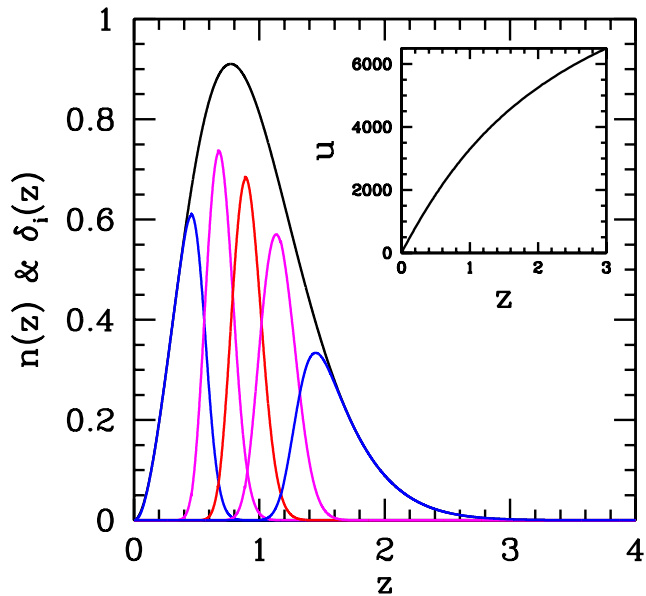}
\end{center}
\vskip -.7truecm
\caption{Distribution of galaxy redshift values and actual shape of
  (5) bins, if defined by using photometric redshift. In the inner
  frame the $z$--$u$ relation is shown ($u$ in Mpc). The plots are for
  a spatially flat model with $\Omega_m = 0.24$ and $H_0=73~$km/s/Mpc.}
\label{EN5}
\end{figure}
\begin{figure}
\begin{center}
\vskip -.4truecm
\includegraphics[scale=0.45]{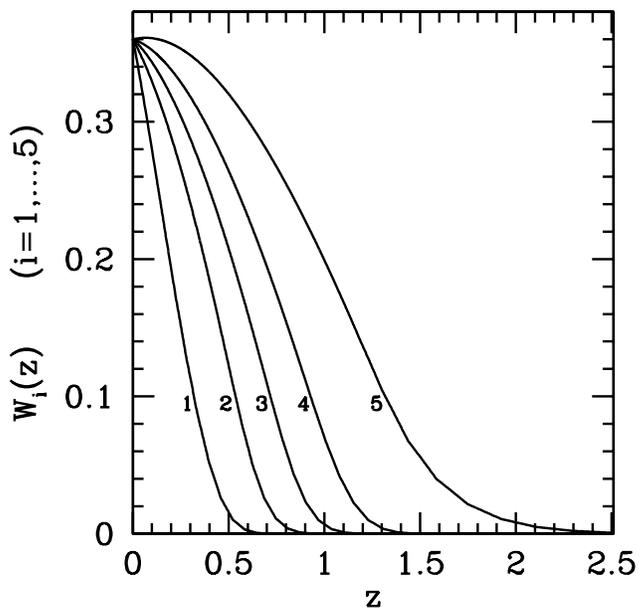}
\end{center}
\vskip -.6truecm
\caption{$W_i(z)$ functions, yielding the effective distribution of
  lensing systems, in the 5--bin case. }
\label{wiz}
\end{figure}
yielding the distributions
\begin{equation}
\delta_{i}(z) = {D_{i}(z) \over \int_{0}^{\infty}D_{i}(z')dz'}
\end{equation}
as a function of $z$ of the actual setting of lensed objects. Figure
\ref{EN5} exhibiting the resulting redshift bins, shows that
increasing their number causes extensive overlaps, which risk
  polluting the analysis when the number of galaxies per bin is too
  small.

Within Figure \ref{EN5} we also show the conversion between $z$ and
$u$ (in Mpc), for a specific model consistent with WMAP-7 CMB results
\cite{wmap7}, with $\Omega_m = 0.24$ and $H_0=73~$km/s/Mpc
(matter density and Hubble parameter). This model will be used
throughout this work, in order to exploit the results of wide
simulations with a large dynamical range, including also baryon
physics, available to us.

The images of any galaxy belonging to a bin can be lensed by systems
at lower $z$. From the distributions $\delta_i(z)$ we therefore derive
the functions
\begin{equation}
  F_i (z) = \int_{\Delta z_i} dz'~\delta_i(z') \left[1-{u(z) \over u(z')} 
  \right]\, \, .
\end{equation}
The factor in square brackets must be set to zero if negative: the
lens is closer than the lensed galaxy. Such functions then yield the
window functions
\begin{equation}
\label{WI}
W_i (z) = {3 \over 2} \Omega_m F_i(z) (1+z)~
\end{equation}
to be used in eq.~(\ref{pijl}). In Figure~\ref{wiz} we show the $W_i(z)$
profiles in the 5--bin case.
\begin{figure}
\begin{center}
\vskip -.5truecm
\includegraphics[height=8.cm,angle=0]{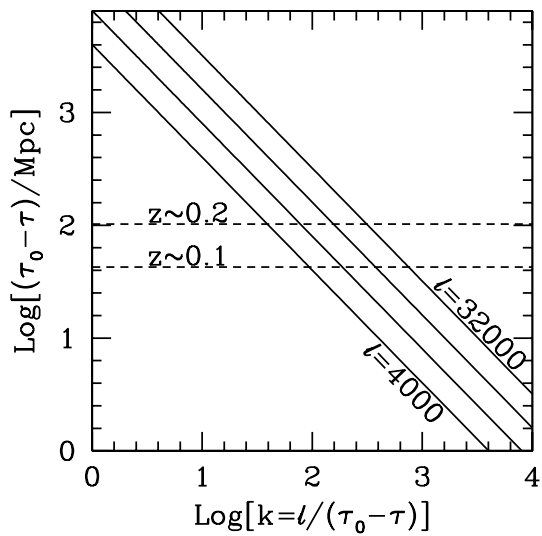}
\end{center}
\vskip -.6truecm
\caption{To obtain shear spectra, the fluctuation spectra $P(k,u)$ are
  to be integrated. The lines are examples of the domains of
  integration, for logarithmically equispaced $\ell = 4000,$ 8000,
  ....., 32000~, on the plane spun by $k$ and $u=\tau_o-\tau$. }
\label{kl}
\end{figure}

\subsection{Fluctuation spectra}
The dependence of $C_{ij}$ on $\ell$ can be roughly interpreted as a
dependence on an angular aperture $\vartheta \sim 2\pi/\ell$ which
subtends linear scales $\lambda = \theta\, u$, increasing with $z$. In
turn, the rough correspondence $k \sim 2\pi / \lambda$ also
holds. Altogether, a given $\ell$ corresponds to decreasing $k$ values
as $u$ (or $z$) increases.

Accordingly, in eqs. (\ref{pijl}), the fluctuation spectrum is taken
at decreasing $k$ values, as $u$ increases. In principle, for $u \to
0$, $P(k)$ should be evaluated at $k \to \infty$ where, however, it
vanishes.

If we consider the $\log k$--$\log u$ plane, the integration in
eq. (\ref{pijl}) is carried along tilted straight lines such as those
shown in Figure \ref{kl}, each line corresponding to a given $\ell$.

The spectra $P(k,z)$ therefore yield functions $P_\ell(u)$, where
$\ell$ fixes a line on the $\log k$--$\log u$ plane and $u$ is used as
an abscissa for such a line. More explicitly, it will be $P_\ell (u) =
P[\ell/u,z(u)]$.

\subsection{From fluctuation to shear spectra}
The integration interval in eq.~(\ref{pijl}) is apparently finite.
When $u$ reaches the conformal age of the Universe $\tau_0$, however,
$z$ approaches $\infty$. Figure \ref{wiz} shows that all $W_i$ vanish
well before so, and this sets an effective upper limit to the
integration.

Aiming at 6--digit precision, a numerical integration performed by
summing on a large number of equispaced points requires $\sim 5000$
points up to $u \sim 6000$ ($z \sim 3$). By itself, however, a
  large number of points does not guarantee a safe result; what
  matters is that $P_\ell(u)$ has a fair value at each $u$
  considered. The point is that $P_\ell(u)$, for each $\ell$, might
  result from interpolating along a set of points $u_r$ and the key to
  obtain fair $C_{ij}(\ell)$ values is that the number of $u_r$ is
  adequate.

Two cases are considered here: we first used Halofit \cite{smith},
enabling us to evaluate approximated spectra at any $k=l/u$ for any
$u(z)$; then, we used more precise spectra worked out from
simulations, although at a given set of redshifts.

In principle, in the former case, the Halofit package can be directly
questioned for each $k$ and $u$. {\it Vice-versa,} Halofit can also be
exploited to test how many $k_s$ and $u_r$ values are needed to obtain
interpolated results equivalent to those resulting from ``direct
questioning''.

A point one appreciates soon is that interpolating along the tilted
lines of Figure \ref{kl} is not so effective as interpolating among
spectra at constant $k$ along $u$ (or $z$). More quantitatively, {
  Halofit} was used to test that using the results of a large
simulation, whose fluctuation spectra are known at the redshifts
\begin{equation}
1+z_r = 10^{r/20} \, \, \, \, \, \, \, (r=1,\, ....\, ,19)\, \, 
\label{zr}
\end{equation}
is adequate to achieve the required precision. This is true if we
interpolate at constant $k$. In contrast, if we deduce first
$P_\ell(u_r)$ values and interpolate then among them, fluctuation
spectra at more $z_r$ (approximately 3 times as many) are needed.  }

The cosmology for which we consider Halofit spectra is the same used to
run the simulation in the next Section. More specifically, we take a
flat $\Lambda$CDM model with $\Omega_m = 0.24$, $\Omega_b= 4.13 \times
10^{-2}$, $h=0.73$, $n_s = 0.96$, (density parameters of total matter
and baryons, Hubble parameter, primordial spectral index,
respectively) and normalized so that the m.s.a. of density
fluctuations, at $8\, h^{-1}$Mpc, $\sigma_8 = 0.8$ at $z=0~.$

For our purposes it is necessary to make also use of Gaussian
integration, i.e. to project the integrand function $f(x)$ onto
polynomials $\pi_\alpha(x)$, orthogonal with an assigned weight
function $R$$(u)$, finding its components $f_\alpha$. The integrals
$\Pi_\alpha$ of each polynomial are then known and $\sum_{\alpha=1}^N
f_\alpha \Pi_\alpha$ is a reliable integral of $f(x)$, if $N$ is large
enough. As is known, this technique can be translated into a practical
and simple procedure, so that integration is reduced to a weighted sum
of values taken by the integrand function $f(x)$ in a suitable set of
points~$x_\alpha$.

More in detail, using {\it monic} polynomials, we have that
\begin{equation}
  \int_0^\infty dx \, \, R(x) \, \pi_\alpha(x) \pi_\beta(x) = {\cal N}
  \delta_{\alpha\beta} \, \, ,
\end{equation}
with a known normalization $\cal N$. Monic polynomials are obtained
from a suitable recurrence relation assuming that the coefficient of
the leading term, for each $\alpha$, is unity. If we then truncate the
sum to $N$ terms, the $N$ zero's of $\pi_N (x)$ are the points
$x_\alpha$, while the corresponding weights are
\begin{equation}
w_\alpha = {\int_0^\infty dx\, \, R(x) \pi_{N-1}^2(x)
\over \pi_{N-1}(x_\alpha) \pi'_{N}(x_\alpha)} \, \, \, ,
\label{wr}
\end{equation}
$\pi' (x)$ being the ordinary derivative of $\pi (x)$. Then,
\begin{equation}
\int_0^\infty dx\, f(x) = \sum_\alpha w_\alpha f(x_\alpha)
\label{sum}
\end{equation}
and, for $R(x) \propto e^{-x}$, this technique is dubbed Gauss--Laguerre
integration, as $\pi_\alpha (x) = L_\alpha (x)$, the Laguerre
polynomials.

In the case of eq. (\ref{pijl}), where integration is cut off by the
exponential--like decay of the $W_i$ functions, this approach can be
applied by assuming that $u = \phi(x)$ or, more specifically, $x =
(u/\bar u)^\beta$ and suitably selecting then $\bar u$ and $\beta$,
  
We shall show the degree of approximation allowed by such a technique
after discussing, in the next subsection, why it is needed and which
are the limitations to~$N$.

\subsection{Formal inversion}
We then rewrite eq. (\ref{pijl}) as follows:
\begin{equation}
\label{linear1}
c_A(\ell) = \sum_{r=1}^N w_r S_{A,x_r} p_{x_r} (\ell) \equiv
\sum_{r=1}^N {\cal M}_{Ar}  p_{r} (\ell)\, \, \, .
\end{equation}
Here we have set
\begin{equation}
A \equiv ij\, ,\, \, \, \, 
c_A = C_{ij}/H_0^4
\end{equation}
with the correspondence law
$$
\matrix{
i,j & 1,1 & ... & 1,5 & 2,2 & ... & 2,5 & 3,3 & ...  & 5,5 \cr
A & 1 & ... & 5 & 6 & ... & 9 & 10 &  ... & 15
}
$$
while
\begin{equation}
S_{A,x_r} = W_i(x_r) W_j(x_r)/R(x_r)\, ,
\end{equation}
\begin{equation}
p_{r} (\ell) \equiv p_{x_r} (\ell)= 
P_\ell[\phi(x_r)] = P[\ell/\phi(x_r),\phi(x_r)] \,  .
\end{equation}
If, in eq. (\ref{linear1}), we take $N=15$, $ {\cal M}_{Ar} $ are
square matrices and, provided that they are not singular, the inverse
equation
\begin{equation}
 p_{x_r} (\ell) = \sum_A ({\cal M})^{-1}_{rA} c_A(\ell)
\label{solution}
\end{equation}
also holds. Accordingly, we shall be able to recover the spectrum
$P(k,z)$ for any $k = \ell/\phi(x_r)$, at the redshift values
$z[\phi(x_r)] $.

Notice that the inversion procedure acts on each $\ell$ value
separately. Any $z$ and/or $k$ value can be attained, in principle,
just by suitably choosing the $\bar u$ and $\beta$ parameter, to
obtain a suitable $\phi(x)$ function. In principle, the choice can
depend on $\ell$.

All this makes it clear that Gauss--Laguerre summations
(\ref{linear1}), in this case, must be limited to $N=15$. This is a
consequence of using 5 bins. With 3 bins, $N=6$ at most, a value
inadequate to yield any reliable integration. On the contrary if,
  e.g., we take 7 (10) bins, we have $N = 28$ (55). When increasing
  the number of bins, we therefore approach an increasingly
  satisfactory situation, as many terms can be set into the summation
  (\ref{linear1}) and, accordingly, many more linear equations can be
  used.

In order to divide the galaxy sample into many redshift bins, however,
we must have either reliable redshift values for most lensed galaxies,
or a sample including very many galaxies. The first option probably
requires one to do better than using photometric redshifts; the latter
option is the one pursued by Euclid. In this paper, however, we shall
consider only the 5--bin case.

\section{Operational problems}
When trying to exploit this formal inversion we find two kinds of
difficulties: (i) the performance of the integration procedure; (ii) a
quasi--singular behavior of the matrix~${\cal M}_{Ar}$.

Both of them can be, at least partially, overcome and the results we
give here aim to show that the procedure is effective.

\subsection{Integrations}
For instance, in order to avoid a $ {\cal M}_{Ar}$ singular behavior,
a possible option is to reduce its dimension. Before discussing this
in more detail, we discuss the results of a Gauss--Laguerre
integration when reduced to 12 points.

\begin{figure}
\begin{center}
\vskip -.5truecm
\includegraphics[height=11.6cm,angle=0]{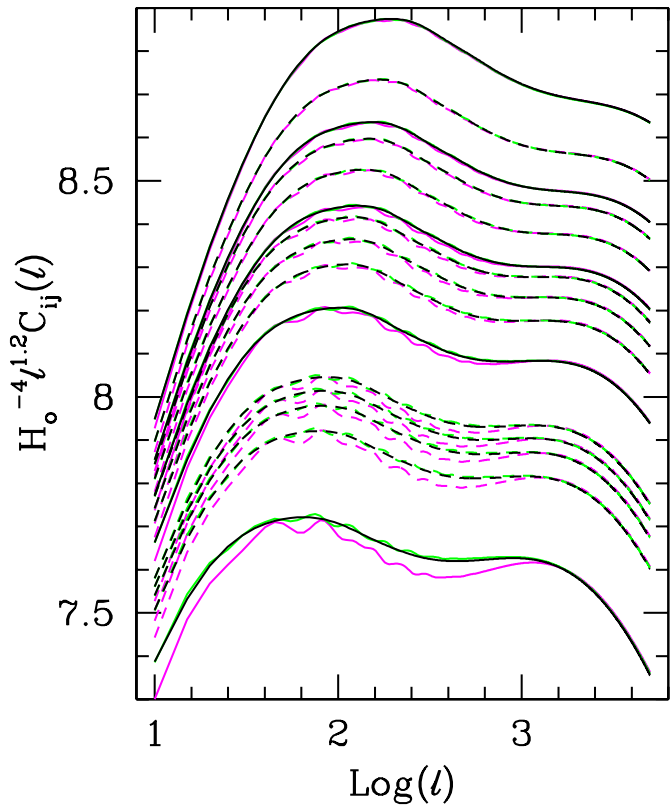}
\end{center}
\vskip -.3truecm
\caption{Tomographic shear spectra obtained from Halofit fluctuation
  spectra. We plot $C_{ij} \times \ell^{1.2}$, so to reduce the
  ordinate range and stress spectral discrepancies. Black spectra from
  Riemann integration (10000 points). Green and magenta spectra from
  the Gauss--Laguerre integration with different choices of $\bar u$ and
  $\beta$ (see text). Values $\ell = 10,\, \, 15, \, 20, \, .... \,
  ,\, \, 5000\, $ are plotted. Spectra are larger for greater $i,\,
  j$. Solid lines are $C_{ii}$ ($i=1,...,5$). Dashed lines are
  $C_{ij}$ ($j>i$). Notice the gradual decrease of discrepancies
  towards greater $i,\, j$, due to a wider $z$--range contributing
  to integration, i.e., to a richer data system. }
\label{integral}
\end{figure}

In Figure \ref{integral} we show $\ell^{1.2} c_A(\ell)$ ($A = 1,\,
...\, 15$) obtained by using Halofit spectral expressions.
$\ell$ values from 10 to 5000, at intervals of 5, are taken. Three
integration techniques are compared: (i) The black curve is the
benchmark obtained by performing a Riemann integration with 10000
points between $u=0$ and $6000\, \, $Mpc. (ii) The magenta curve is
the Gauss--Laguerre integral with $\bar u = 1200$ and $\beta =
2.9$. (iii) The green curve is also the Gauss--Laguerre integral with
$\bar u = 600$ and $\beta = 1.9\, .$

This Figure allows us to draw some immediate conclusions: (a) The
performance of a Gauss--Laguerre integration exhibits a strong
dependence on the parameter choice. (b) Results are however better for
larger $i$, $j$.

The reason for point (b) is soon evident: The redshifts $z(u_r)=
z[\phi(x_r)]$, shown in Table I, are distributed between 0 and $\sim
1.5$. All of them yield a substantial contribution to the sum for
large $A$. On the contrary, owing to the fast cut off of $W_i$, for
small $i$ values, at low $A$ the sum risks including only a few terms
yielding a real contribution.
\noindent
This makes it also clear that using $u$ values corresponding to $z >
\sim 1.5$ is vain. The contribution to the integrals coming from
larger $z$ are to be approximated by the assumed exponential decay of
the integrand function and the shape of the orthogonal polynomials
selected.

It is then also clear why the (iii) integration procedure yields
better results than the (ii) one: In the former case, the $u_r =
\phi(x_r)$ values are more densely accumulated at lower $z$, thus
allowing better integration for low $A$. Apparently, this causes no
detriment to high--$A$ results.

\vskip .3truecm
\centerline{Table I}
\vglue .1truecm
\centerline{$z_r$ values}
\vglue -.05truecm
$$
\matrix{ 
\hline
\cr
{\rm (ii)\, \, case }
\cr
 .0466 & .1152 & .1886 & .2664 & .3499 & .4402 \cr .5386 & .6473
  & .7700 & .9122 & 1.0850 & 1.3162 \cr 
\cr
\hline
{\bf }
\cr
{\rm (iii)\, \, case}
\cr
.1234 & .2338 & .3338 & .4304
  & .5263 & .6237 \cr .7245 & .8307 & .9450 & 1.0714 & 1.2171 & 1.4004
\cr
{\bf }
\cr
\hline
{\bf }
}$$
\vskip .3truecm

Of course, small shifts of $\beta$ or $\bar u$ cause no substantial
difference. However, although exploring different options with
much care, we cannot exclude that better $\bar u$, $\beta$ choices
exist. In particular, if one uses a different measure $R(x)$ on the
integration interval, different orthogonal polynomials follow. For
instance, one could use $R(x) \propto e^{-x^2}$ yielding $\pi_\alpha(x)
= H_\alpha(x)$, the Hermite polynomials; or some other $R(x)$ yielding
non--tabulated polynomials.

\subsection{The SVD technique}
Let us then consider the inversion procedure. The problem here is that
the matrix 
\begin{equation}
{\cal M}_{Ar} = w_r W_i(x_r) W_j(x_r)/R(x_r)
\end{equation}
tends to be singular. There is a specific analytical reason for that:
when $u_r = x_r^{1/\beta} \bar u$ exceeds $\sim 0.7$ any matrix
element containing $W_1$ tends to vanish, as is evident from Figure
\ref{wiz}. A similar feature is caused by any element containing a
generic $W_i$ with $i \neq 5$, when $u_r$ yields a redshift $z \sim
1.5~.$ But, even keeping all $u_r$ below 1.5, the ratio between
largest and smallest diagonal elements tends to be too large, even in
double precision.

Before further discussing this point, let us introduce a specific
technique, allowing us to gauge the degree of singularity of a matrix,
dubbed {\it SVD} (singular value decomposition).

It is based on a theorem of linear algebra, stating that any real $N_r
\otimes N_c$ matrix $\cal M$, with $N_r \geq N_c$, can be decomposed
into a rows $\times$ columns product 
\begin{equation}
{\cal M} = {\cal U} \times \left|~diag(s_i)~\right| \times {\cal V}^T.
\end{equation}
Here ${\cal V}^T$ is the transpose matrix of a matrix ${\cal V}$
which, as well as ${\cal U}$, is orthonormal, while {\bf s} is
diagonal.  Apart from multiplicative factors, the decomposition is
unique. If $N_r = N_c$, the inverse of $\cal M$, in general, reads
\begin{equation}
{\cal M}^{-1} = {\cal V} \times \left|~diag(1/s_i)~\right| \times {\cal
  U}^T~,
\end{equation}
and the technique is also a valid numerical way to invert large
matrices. 

The degree of singularity, however, can be inspected by just
considering the $s_i$ components. One (or more) vanishing element(s)
cause the matrix to be singular. Even if it is not so, however, and
the ratio between the greatest and smallest $s_i$ exceeds $\sim 10^6$
($10^{12}$), there is no hope to invert ${\cal M}$ in {\it single}
({\it double}) precision. Anyhow, if a level of precision $\cal
O$$(1:10^6)$ is to be kept, one must use {\it double} precision
keeping the highest $s_i/s_j$ ratio within $\sim 10^6$.

A fair discussion of the technique can be found in {\it Numerical
  Recipes} \cite{NR} or in {\it Matrix Computations} \cite{golub},
where is also discussed how this technique can be applied when $N_r >
N_c$, as well as what to do to find tentative solutions when the
highest $s_i/s_j$ ratio is too large, and even when some $s_i$
vanishes.

Here we shall not debate this approach any further. We shall just
report that it allowed us to test a wide number of options, by
selecting those exhibiting a lower level of singularity. Although our
inspection was systematic and detailed, we cannot exclude that even
more efficient solutions can be found. Here we wish to outline that
solutions are indeed available, and then discussing which problems
remain.

\subsection{Inversion}
Making use of 12 Gaussian points to integrate, eq.~(15) can then be
considered as a linear system of 15 equations with 12 unknowns
$p_r(\ell)$ ($r=1,..,12$). The very $SVD$ technique is built to deal
with such cases, and the very redundancy of the system is a reason
for keeping the top $s_i/s_j$ value within $\sim 10^6$.

\begin{figure}
\begin{center}
\vskip -.6truecm
\includegraphics[height=9.cm,angle=0]{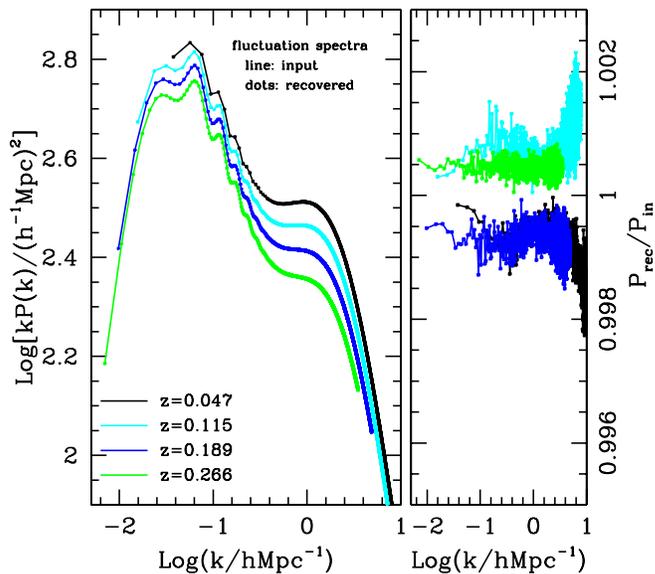}
\end{center}
\vskip -.7truecm
\caption{Recovery of $P(k,z_r)$ spectra from $C_{ij}(\ell)$; the
  values $z_r$ are yielded by $x_r$ values. Here the 4 lowest $z$
  values are shown. The ratio of input to recovered values never
  differs from unity more than $\simeq 10^{-4}$.}
\label{invert}
\end{figure}

\begin{figure}
\begin{center}
\vskip -.6truecm
\includegraphics[height=9.cm,angle=0]{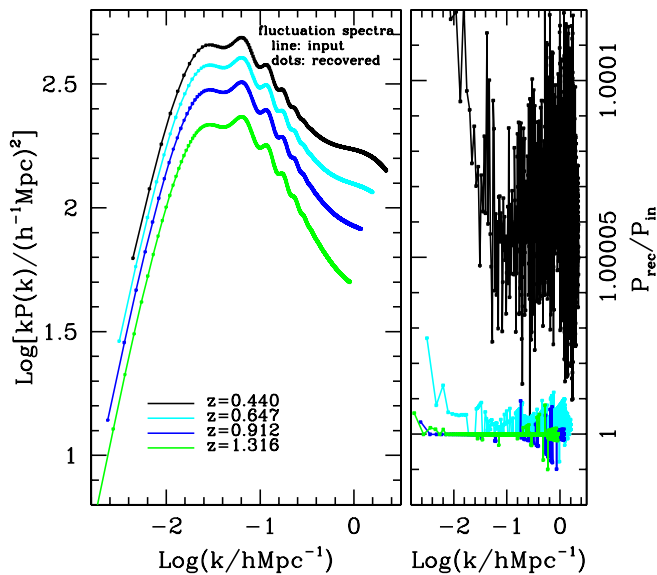}
\end{center}
\vskip -.7truecm
\caption{As the previous Figure, for $z_r$ with $r=6,8,10,12 $.
  Discrepancies, for larger $z_r$ are even smaller,
  keeping well within 1:$10^6$ for $r>8$.  }
\label{invert1}
\end{figure}

If we try to recover the $P(k,z)$ spectrum from the spectra
$C_{ij}(\ell)$, we then have full success. This is shown in Figures
\ref{invert} and \ref{invert1}.
The degree of precision is better for higher redshift values: we pass
from $\sim 0.2\, \%$ at low $z$, to a precision $\cal O$$(1:100000)$
for $z = 1.316$, the highest redshift considered.

A point to outline is also the shift of the $k$ interval explored when
$z$ increases. The excellent results for the highest redshift, e.g., are
concentrated within an interval where non linearity is just
approached, also because the non--linear $k$--range shifts to the
right at higher redshift. With Halofit spectra, however, one can hardly
do better.

We tested the $SVD$ algorithm also using 12 equations, excluding
either the 3 top--$A$ equation, or the equations for $i=1$,
$j=1,2,3$. The inversion succeeds, discrepancies keep an acceptable
level, but are indeed larger (and somewhere notably larger) than those
obtained with the full set of equations. In the former case, the worst
output/input ratio is at intermediate redshifts, where errors increase
up to a factor 20~. In the latter case, the worst results concern
highest redshift values, where the errors have a really significant
increase, by a factor up to $\sim 4 \times 10^4$, although however
keeping within $\cal O$$(5 \, \%)$.

Figures \ref{invert} and \ref{invert1} are one of the main results of
this work. To operate the inversion, we made use of the very
$C_{ij}(\ell)$ obtained by using a Gaussian integration procedure
(black curves in Figure \ref{integral}). In turn, Gaussian--Laguerre
integrals were based on spectra obtained from interpolating Halofit
spectra at the redshifts $1+z_r = 10^{r/20}$ ($r=0,1,...,19$), in
order to retain homogeneity with the results obtainable from
simulations (see next Section).  Interpolation was performed at
constant $k$ values.

The next point we whall inspect here is the ``stability'' of the
inversion procedure. We do so by applying the inversion algorithm to
the angular spectrum worked out by performing a Riemann integration.

\begin{figure}
\begin{center}
\vskip -.6truecm
\includegraphics[height=13.cm,angle=0]{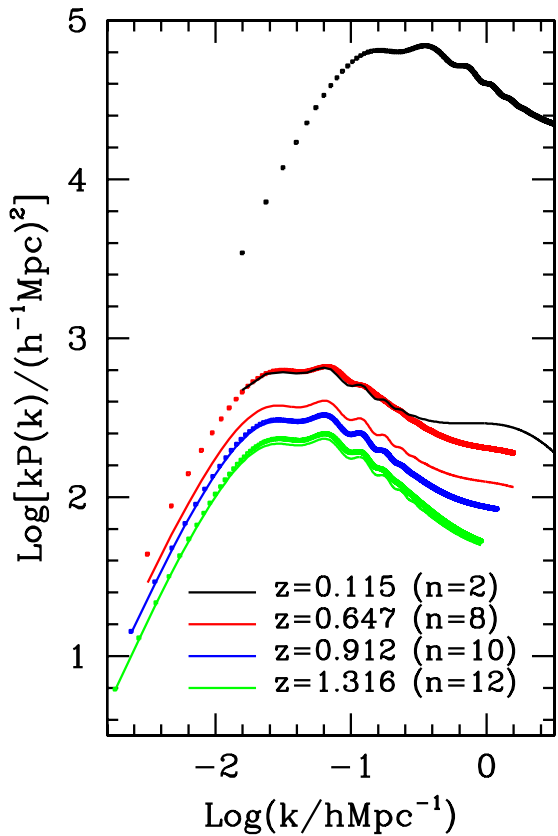}
\end{center}
\vskip -.7truecm
\caption{Results of the inversion algorithm based on 12 point
  integration, if applied to ``exact'' spectra. Although the power
  spectrum is better approximated at high redshift, the result is
  unsatisfactory, namely because the $z$ dependence of the power
  spectrum is not reproduced. }
\label{badinvert}
\end{figure}

This operation does not yield an immediate success. The discrepancies
between $C_{ij}(\ell)$ worked out with 12 Gaussian points or 10000
Riemann points is fairly large in some points, up to $\cal O$$(10\,
\%)$ and in Figure \ref{badinvert} we show the results of a brute
inversion for large $z_r$. The results of this procedure are better at
large redshifts, where the discrepancy between $C_{ij}(\ell)$ obtained
through Riemann and Gauss integrations is smaller. 

It is however clear that the procedure is to be tested when using more
than 12 Gaussian points, so reducing the discrepancy between Gaussian
and Riemann integrations well below the expected noise level.  Rather
than the Gaussian--Riemann discrepancies, it will then become
important to test the impact of random noise.

However, still with 12 Gaussian points, there is a possible
improvement of the results, that we shall debate after discussing the
replacement of Halofit spectra with simulation spectra.

\section{Beyond Halofit}
In this Section we will report the results of the same procedures,
when applied to spectra obtained from large hydrodynamical
simulations. This enables us to consider larger $\ell$ values and
shear spectra exploring scales well inside galaxy clusters, which
cannot be approached if baryon physics is disregarded. We also recall
that several authors \cite{chh,semboloni} discussed the limitations in
the Halofit reconstruction of non--linear power spectra, whose use was
dangerously extended to models including DE with an equation of state
different from $w \equiv -1$.

\subsection{Simulation}
The simulation used here follows the development of structures within
a periodic box of comoving side $L = 410\, h^{-1}$Mpc, where $(2
\times) 1024^3$ particles are set. The cosmology is a spatially flat
Gaussian $\Lambda$CDM model with $\Omega_m = 0.24$, $\Omega_b= 4.13
\times 10^{-2}$, $h=0.73$, $n_s = 0.96$. Fluctuations are normalized
so that the m.s.a. of density fluctuations, on the scale of $8\,
h^{-1}$Mpc, $\sigma_8 = 0.8$ at $z=0~.$ The two populations of
$1024^3$ particles, therefore, have masses $m_c \simeq 1.89 \times
10^9\, h^{-1} M_\odot$ and $m_b \simeq 3.93 \times 10^8\, h^{-1}
M_\odot$.

The simulation was carried out by using the TreePM-SPH GADGET-3 code
(SPH: smoothed--particle hydrodynamics), an improved version of the
GADGET-2 code (Springel 2005). Initial Zeldovich displacements were
generated at $z_{in} = 41$. Gravitational forces were computed using a
Plummer--equivalent softening which is fixed to a physical scale
$\epsilon_{Pl}=7.5\, h^{-1}$kpc from $z=0$ to $z=2$, being then
constant in comoving units at higher redshifts.

As far as baryon physics is concerned, radiative cooling was computed
for non--vanishing metallicity according to Sutherland \& Dopita
\cite{suth}, also including heating/cooling from a spatially uniform
and evolving UV background. Gas particles above a given threshold
density are treated as multi-phase, so as to provide a sub–resolution
description of the inter–stellar medium, according to the model
described in ref.~\cite{spring}. Within each multi-phase gas particle,
a cold and a hot-phase coexist in pressure equilibrium, with the cold
phase providing the reservoir for star formation. Conversion of
collisional gas particles into collisionless star particles proceeds
in a stochastic way, with gas particles spawning a maximum of two
generations of star particles. The simulation also includes a
description of metal production from chemical enrichment contributed
by SN-II and SN-Ia supernovae and asymptotic giant branch (AGB) stars,
as described in ref.~\cite{torna}.  Stars of different mass,
distributed according to a Salpeter IMF, release metals over the
time-scale determined by the corresponding mass-dependent
life-times. Kinetic feedback is implemented by mimicking galactic
ejecta powered by SN explosions. In these runs, galactic winds have a
mass upload proportional to the local star-formation rate. The wind
velocity is then $v_w = 500\, {\rm km/s}$; this corresponds to
assuming about unit efficiency for the conversion of energy released
by SN-II into kinetic energy for a Salpeter IMF. More detail on this
simulation are provided in \cite{casarini2012}.

Simulations including baryon physics, aimed at evaluating shear
spectra, were performed by various authors (see, e.g.,
\cite{casarini2011, semboloni}). Van Daalen et al.~\cite{vandaalen}
included also AGN feedback in their simulations. The simulation used
here, however, was performed in a box and with a dynamical range large
enough to enable us to evaluate fluctuation spectra for the whole
range needed to compute $C_{ij}(\ell)$ up to $\ell \simeq
40,000$--50,000~.

Simulations including baryon physics, aimed at evaluating shear
spectra, were performed by various authors (see, e.g.,
\cite{casarini2011, semboloni}). Van Daalen et al.~\cite{vandaalen}
included also AGN feedback in their simulations. The simulation used
here, however, owns two suitable features: (i) it was performed in a
box large enough ($410\, h^{-1}$Mpc) to allow us a direct connection
between simulation and linear spectra; (ii) a dynamical range large
enough was used, enabling us to evaluate $C_{ij}(\ell)$ up to $\ell
\simeq 40,000$--50,000~, by exploiting fluctuation spectra unaffected
by numerical noise.

\subsection{Fluctuation spectra}
Power spectra are computed at the redshift values (\ref{zr}), by using
the algorithm PMpowerM included in the PM package, courtesy of
A. Klypin.

Through a CiC procedure the algorithm assigns the density field on a
uniform Cartesian grid starting from the particle distribution. It
then calculates the spectrum through a FFT on a $n^3$ ($n=2^4N$) grid;
i.e. $n = 1024 \times 2^4 = 16384~.$ These large $n$ are obtainable by
considering a $N^3$ grid in a box of side $L/2^4$, where simulation
particles are inset, in points with coordinates $x_{i,f} = x_i - \nu
L/2^4$ ($i=1,2,3$), and an integer $\nu$ selected so that $0 < x_{i,f}
< L/2^4$. This technique provides spectra down to wavelengths slightly
above the gravitational softening scale, being only limited by the
numerical noise due to the grid used to set the initial
conditions. The limitation would be however identical if large--$n$
grids could be directly applied to the original box of side $L$.

\begin{figure}
\begin{center}
\vskip -.7truecm
\includegraphics[height=11.5cm,angle=0]{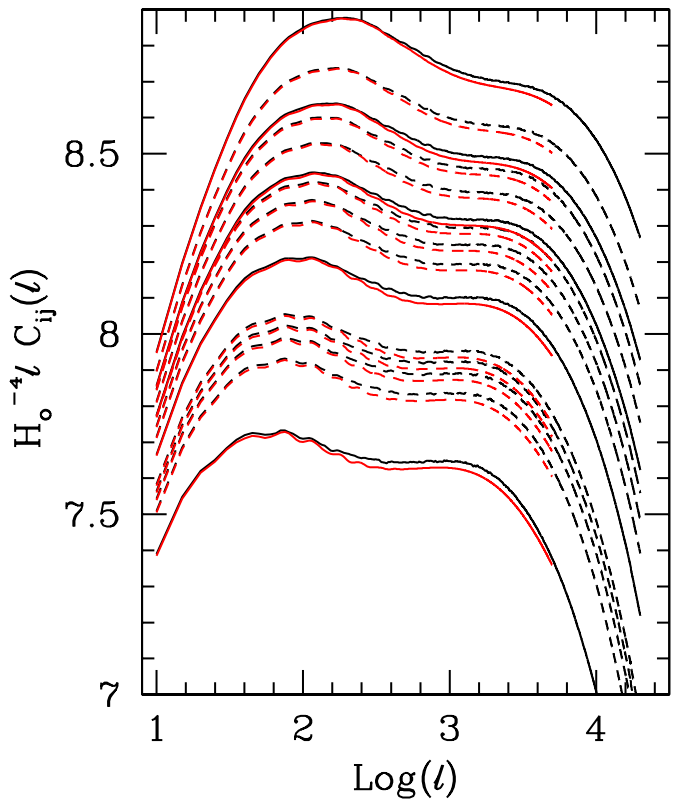}
\end{center}
\vskip -.1truecm
\caption{Tomographic shear spectra for $i,j = 1,\, ....\, ,5$. Solid
  and dashed lines as in Figure 4. Gauss--Laguerre
  integration performed with case (ii) points. Spectra obtained from
  Halofit (red) are overlapped to those obtained from the hydro simulation
  (black). The latter extend well in the non--linear region.
  Discrepancies between the red and black curves are significant in
  the critical region of non--linearity onset and do not reduce
  towards great $i,\, j~.$ }
\label{spec}
\end{figure}

The spectra shown in Figure \ref{spec} are obtained by merging the
linear spectrum of the model with the simulation spectrum. No recourse
to approximate spectral expressions, like Halofit, is needed to
interconnect the two spectral parts. More details on the techniques to
merge the two spectra can be found in \cite{casarini2012} (see also
\cite{jenkins}).

\subsection{Spectra of cosmic shear}
\begin{figure}
\begin{center}
\vskip -.5truecm
\includegraphics[height=9.cm,angle=0]{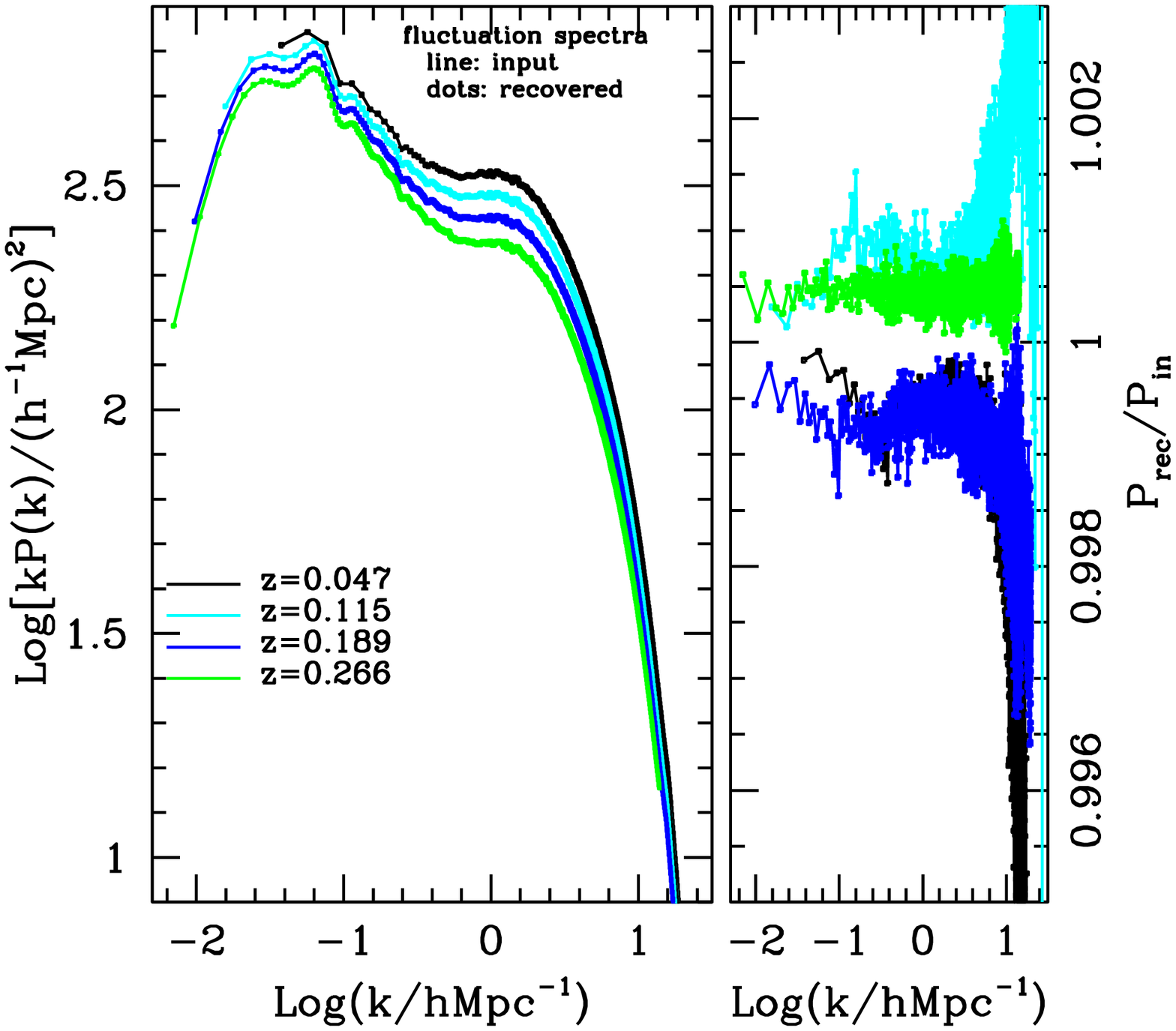}
\end{center}
\vskip -.8truecm
\caption{Recovery of fluctuation spectra from cosmic shear spectra
at the first 4 redshift values use to integrate. This Figure is analogous
to Figure 5, just extending to larger $k$'s.
}
\label{jnvert}
\end{figure}

\begin{figure}
\begin{center}
\vskip -.5truecm
\includegraphics[height=9.cm,angle=0]{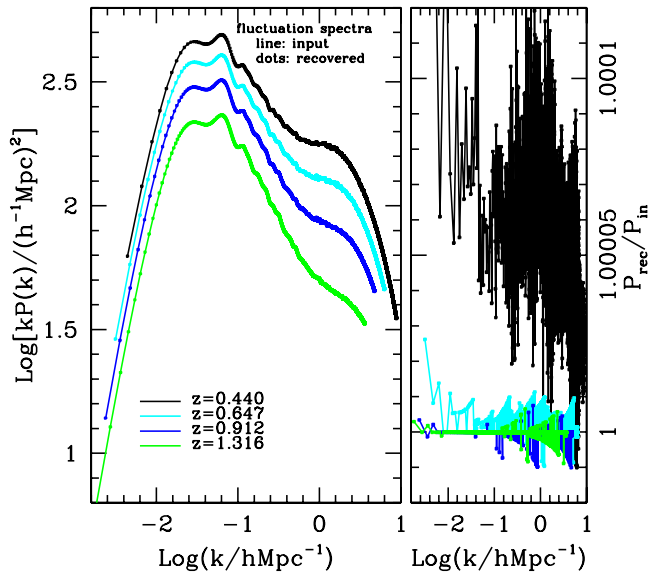}
\end{center}
\vskip -.8truecm
\caption{As previous Figure, for 4 more redshift values. Also here,
as in the Halofit case, discrepancies are smaller for larger $z$ values.}
\label{jnvert1}
\end{figure}

The recovery of fluctuation spectra works on each $\ell$ value
separately: spectra at different $k$ values for each redshift $z$ are
obtained by relating the values taken by the 15 $C_{ij}$ at that
$\ell$. The curves in Figures \ref{jnvert} and \ref{jnvert1},
accordingly, arise from the merging of contributions coming from the
large set of $\ell$ values available. 

Formally, all $C_{ij}$ contribute to recover $P(k,z)$ at any $z$.
However, the physical reason why the recovered $P(k,z)$ is slightly
less precise at small $z$ resides in the fact that the $C_{ij}(\ell)$
are almost--vanishing for the significant $\ell$ values. This is also
the reason why the inversion procedure is more efficient when more
Gaussian points fall at low $z$.

\begin{figure}
\begin{center}
\vskip -.5truecm
\includegraphics[width=11.cm,height=8.5cm,angle=0]{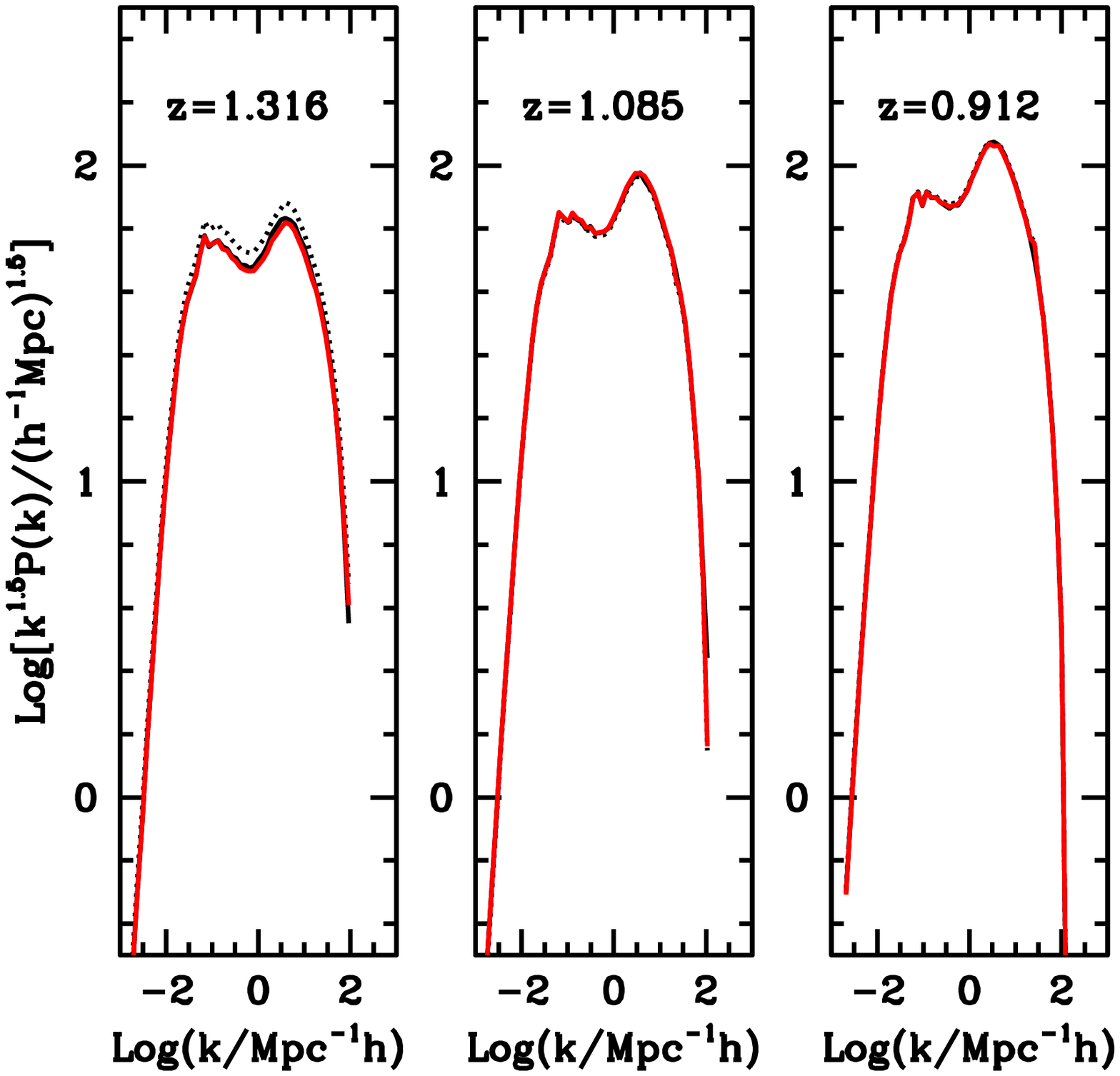}
\vskip -.3truecm
\includegraphics[width=11.cm,height=8.5cm,angle=0]{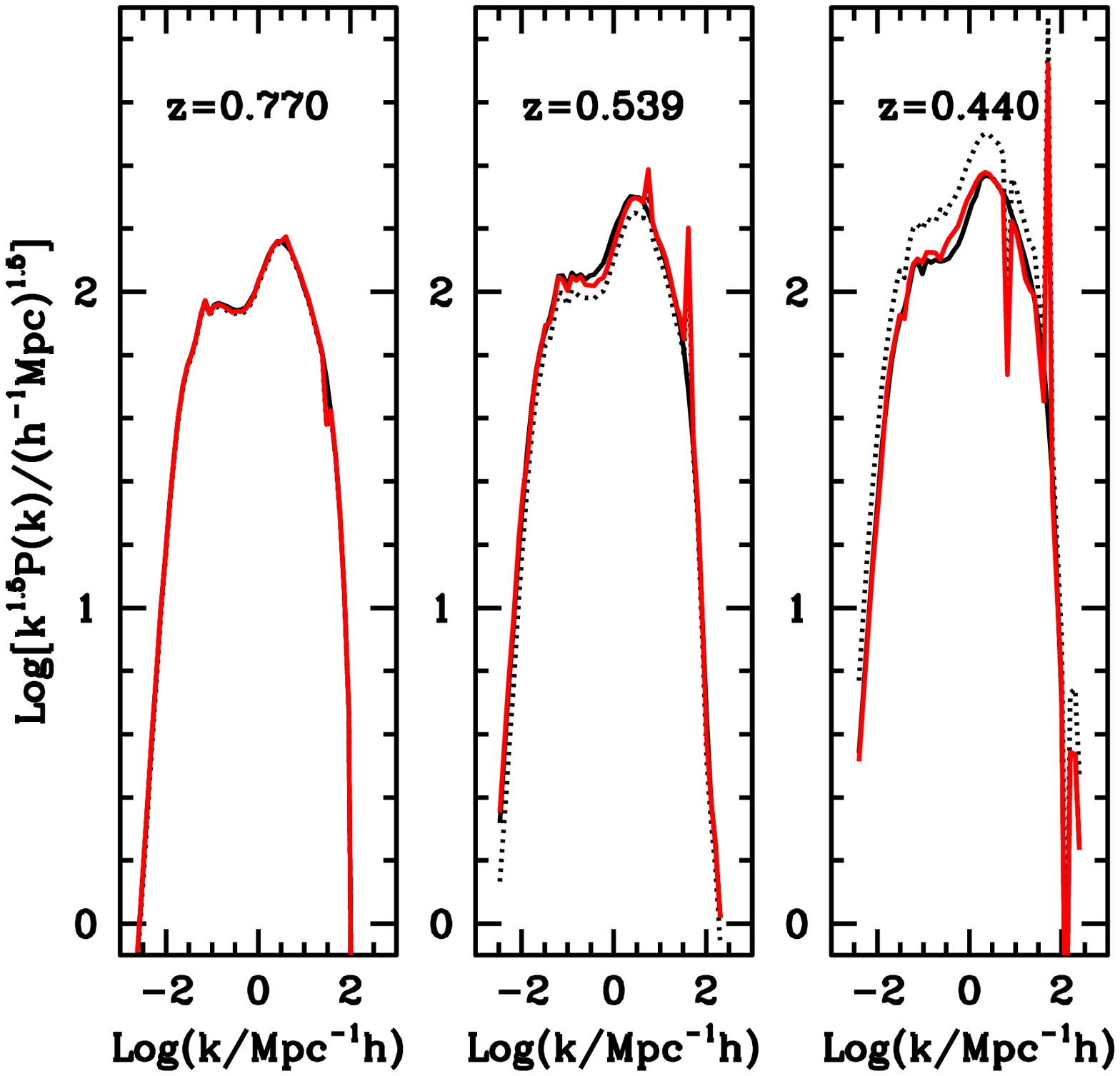}
\end{center}
\vskip -.8truecm
\caption{Results of the inversion algorithm based on 12 point
  integration, if applied to ``exact'' spectra obtained from the
  simulation, by using the renormalization technique described in the
  text. The solid black curves (somewhere invisible) are the input
  fluctuation spectra $P(k,z_s)$ ($z_s$ are the redshifts used by the
  Gaussian 12--point integration algorithm). The dotted curves are the
  spectra obtained by applying the inversion algorithm on the shear
  spectra.  After renormalizing them, we obtain the solid red curves.
  For $z >\sim 0.7$ the discrepancy keeps within 2-3$\, \%$ at all
  $k$'s.  At lower $z$ some spikes appear, indicating a failure of the
  $SVD$ method at specific $\ell$ values. Below $z \sim 0.5$
  discrepancies mostly exceed 10$\, \%$.  }
\label{fairinvert}
\end{figure}

Figure \ref{jnvert} and \ref{jnvert1}, however, show that the relation
(\ref{pijl}) is efficiently invertible, {in principle}, also for
those $\ell$ which deal with the inner cluster structures.

Let us finally return to using the inverse of the matrix $ {\cal M}
_{Ar} $, obtained with 12 Gaussian points, to invert the ``exact''
$C_{ij}(\ell)$ spectra. In Section 3 we already noted that the
question will bear a completely different aspect when we are allowed
to use more than 5 bins. However, even with 5 bins, a substantial
improvement is attained through a suitable renormalization
procedure, essentially requiring the same information needed
to obtain the window functions $W_i(z)$, i.e. the
background ``geometrical'' model parameters.

{\bf If we add to} these parameters {\bf a value of the spectral index
  $n_s$} we easily evaluate the {\it linear} fluctuation spectra
$P^{(c,lin)}(k,z)$ of a pure--CDM model, with the same $\Omega_m$ and
$H_0$. {\bf (Let us however soon outline that the choice of $n_s$ is
  almost arbitrary. In particular we do not need to use the specific
  value used in the simulations.)} A Riemann integration then yields
the angular spectra $C^{(c,lin)}_{ij} (\ell)$ from the fluctuation
spectra $P^{(c,lin)}(k,z)$. In turn, the angular spectra can be
tentatively inverted by using the very $SVD$ ``Gaussian matrix''
worked out for the full model. The final output are linear spectra
$\tilde P^{(c,lin)}(k,z)$ to be compared with the input linear spectra
$P^{(c,lin)}(k,z)$, so building the {\it normalization factors}
\begin{equation}
{\cal N} (k,z) = {\tilde P^{(c,lin)}(k,z) \over P^{(c,lin)}(k,z)}~.
\end{equation}
We shall use these to try to remove the bias, tentatively assuming
that it is the same for the full model and the CDM--linear case.

Let $P(k,z)$ be the spectra obtainable by interpolating the spectra of
the simulation at the redshifts (\ref{zr}). We obtain from them the
shear spectra with either 12--point Gaussian or Riemann
procedures. Then, we calculate the inverted Riemann spectrum $\tilde
P(k,z_s)$, by applying the inverse ``Gaussian matrix'', as above.

These steps give us results not too different from those shown in
Figure \ref{badinvert}. We then renormalize these spectra to obtain
\begin{equation}
\bar P^{R}(k,z) = {\cal N} (k,z) \bar P(k,z)~.
\end{equation}
In Figure \ref{fairinvert} some results of these operations are shown.
{\bf They were obtained by arbitrarily choosing $n_s = 1$. We however
  tested that the residual errors in the renormalized spectra are
  insensitive to the choice of $n_s$ in the interval 0.8--1.15. No
  tests were made outside this interval.  }

At rather large $z$ ($>0.7$), the spectral details can be recovered,
in this way. Therefore, in principle, by re--exploiting the same
information needed to create the window functions, i.e. by assigning
$\Omega_m$ and $h$, we can recover parameters as $\sigma_8$,
$\Omega_b$, $\Omega_c$ (m.s.a. of fluctuations, baryon and CDM density
parameters, respectively), as well as many baryon physics effects
affecting large--$\ell$ shear spectra.  At lower $z$, a precise
spectral recovery is more difficult.

{\bf We wish to add that the procedure still needs to be tested for
  the case when the DM component, or a part of it, is not cold. In
  particular, massive neutrinos, even with a mass in the $10^{-1}$eV
  range, would cause spectral distorsions and might complicate the
  recovery.}

\section{Discussion}
The main achievement of this work is the development of a procedure to
invert the equation yielding shear spectra $C_{ij}(\ell)$ from
fluctuation spectra $P(k,z)$, and its performance tested with spectra
obtained either from Halofit or hydrodynamical simulations.

As is known, lensing data yield the shear spectra $C_{ij}(\ell)$,
while other observables yield the fluctuation spectrum
$P(k,z)$. Converting $C_{ij}(\ell)$ into $P(k,z)$ would then be a
valuable aid to compare the whole dataset with cosmological models.

The stability of the inversion was tested here by applying the
inversion matrix derived from a 12--point Gaussian integration to the
``exact'' $C_{ij}(\ell)$ spectra. The discrepancies between the latter
$C_{ij}(\ell)$ and those obtainable through a 12--point Gaussian
integration were as high as $\sim 10\, \%$, in the worst points.  It
  does not come as a surprise, then, that a brute application of the
  inverse ``Gaussian matrix'' does not yield satisfactory results. A
  renormalization procedure was however introduced, allowing us to
  recover fluctuation spectra for $z >\sim 0.7$ with errors $\sim
  2$--$3\, \%$ and down to $z \sim 0.5$ with errors $\sim 10\,
  \%$. Lower redshift spectra are more difficult to recover.

Such renormalization, requires that a few model parameters, defining
the space--time ``geometry'', which are assigned a priori. Besides
recovering these parameters, the inversion then aims at constraining
the ``dynamical'' parameters, e.g. $\sigma_8$, $\Omega_b$, and, in
addition, the effects of baryon physics.

The restriction to 5 redshift bins and $\sim 12$ Gaussian points
  will be lifted, when Euclid data are available.  With 7 (10)
redshift bins, up to 28 (55) points are usable in Gaussian
integration. In turn, the discrepancy between Gaussian and ``exact''
spectra scales as $(n!)^2/(2n)! \simeq 2^{-2n}$ \cite{AS}. With 20
integration points, it is then $\cal O$$(10^{-6})$ at most. Then this
will no longer be the problem: applying $SVD$ inverted matrix to
``exact'' results or, as matters most, to physical data should be
satisfactory.

The problem to be solved will then concern observational errors. We
expect them to yield a sort of random noise. If so, $SVD$ techniques
provide the best method to deal with noisy matrix inversion. They work
as better as redundancy increases. The point is whether a system of 55
equations dealing with 20--25 unknowns converges to the physical
result, when data have a random noise $\cal O$$(1\, \%)$ or greater.

In our opinion, perspectives are promising; the parameters {\it
    beyond geometry} could be recovered even in the presence of a
  systematic noise 10 times greater than the expected noise in
  data. The main points to test, within this context, are whether the
  technique recovers spectra down to $z \simeq 0$, and how data noise
  affects parameter precision.

In the literature one finds cases when $SVD$ techniques converge, but
to spurious results. This is more and more unlikely as the system
redundancy increases (see, e.g., \cite{golub}). We also recall that,
in our case, the technique is to be applied to each $\ell$ value
separately. It is then even more unlikely that convergence to the same
misleading result separately occurs at any $\ell$. We plan to devote
further work to these issues.

Admittedly, however, Euclid is a somewhat distant perspective. Using
earlier available data, one could however try to achieve a better
inversion, if galaxies could be shared among $\sim 7 $ bins. With the
number of galaxies in datasets soon available, this might become
possible if a substantial improvement in their redshift estimates is
achieved, so that the bin dependence of $C_{ij}$ spectra stands
  out above shot noise. Euclid measures aim to reduce
$\sigma_z/(1+z)$ from $\sim 0.05 $ (the value used here) to $\sim
0.03~$. However, with fewer galaxies one could hope to do better,
perhaps by measuring a spectroscopic redshift for a substantial
fraction of the samples.

Let us finally suggest that the stability problem might can be turned
into an advantage. It is quite likely, in fact, that the inversion
algorithm approaches regular results only when the choice of
``geometrical'' cosmological parameters is correct, yielding otherwise
distorted results, e.g. an awkward $k$ dependence. In association with
the use of $\sim 20$ integration points, this could become a direct
test to recover also {\it background} model parameters.

The route opened by the inversion procedure discussed in this paper
seems therefore promising, and might allow us a more complete
exploitation of data coming from advanced cosmic shear measurements.

\vskip .3truecm

\noindent
ACKNOWLEDGMENTS. SAB acknowledges the support of CIFS though the
contract n.~24/2010 and its extension Prot.n.~2011/338bis~. LC
acknowledges the support of the Brazilian Institutions FAPES and CNPq
and of LUTH-Observatoire de Meudon (France). Thanks are due to
Giuseppe La Vacca for wide discussions. We are grateful to Stefano
Borgani for making available to us his large hydrodynamical
simulations and to Volker Springel for the non-public GADGET-3 code
used to run them. An anonymous referee is also to be thanked for
a number of constructive suggestions.

\vskip .3truecm
\vskip .3truecm


\vskip 2.truecm

\end{document}